\documentclass[aps,pra,twocolumn,amsfonts,floatfix,superscriptaddress,showpacs]{revtex4}

\usepackage{graphicx}
\usepackage{dcolumn}
\usepackage{bm}
\usepackage{mathbbol}

\usepackage{amssymb}          
\usepackage{graphicx}         
\usepackage[latin1]{inputenc} 
\usepackage{amsmath}          
\usepackage{color}

\newtheorem{prop}{Proposition}
\newtheorem{defin}{Definition}
\newtheorem{thm}{Theorem}

\newtheorem{prob}{Problem}

\newcommand{\proof}{\noindent {\bf Proof. }}

\newcommand{\vv}[1]{``#1''}

\newcommand{\ket}[1]{|#1\rangle}
\newcommand{\bra}[1]{\langle #1|}

\newcommand{\Hi}{\mathcal{H}}
\newcommand{\Si}{\mathcal{S}}
\newcommand{\Ei}{\mathcal{E}}
\newcommand{\Ai}{\mathcal{A}}
\newcommand{\Tr}{\mathrm{{Tr}}}

\newcommand{\beq}{\begin{equation}}
\newcommand{\eeq}{\end{equation}}
\newcommand{\bea}{\begin{eqnarray}}
\newcommand{\eea}{\end{eqnarray}}

\newcommand{\cvd}{\hfill $\Box$ \vskip 2ex}
\newcommand{\A}{\mathcal{A}}
\newcommand{\Bi}{\mathcal{E}}

\newcommand{\spec}{\textrm{sp}}
\newcommand{\tr}{\textrm{trace}}

\begin{document}

\title{Single-bit Feedback and Quantum Dynamical Decoupling}

\author{Francesco Ticozzi}
\email{ticozzi@dei.unipd.it}
\affiliation{Dipartimento di Ingegneria dell'Informazione,
Universit\`a di Padova, via Gradenigo 6/B, 35131 Padova, Italy}
\affiliation{\mbox{Department of Physics and Astronomy,
Dartmouth College, 6127 Wilder Laboratory, Hanover, NH 03755, USA}}
\author{Lorenza Viola}
\email{lorenza.viola@dartmouth.edu}
\affiliation{\mbox{Department of Physics and Astronomy,
Dartmouth College, 6127 Wilder Laboratory, Hanover, NH 03755, USA}}

\date{\today}

\begin{abstract}
Synthesizing an effective identity evolution in a target system
subjected to unwanted unitary or non-unitary dynamics is a fundamental
task for both quantum control and quantum information processing
applications.  Here, we investigate how single-bit, discrete-time
feedback capabilities may be exploited to enact or to enhance quantum
procedures for effectively suppressing unwanted dynamics in a
finite-dimensional open quantum system. An explicit characterization
of the joint unitary propagators correctable by a single-bit feedback
strategy for arbitrary evolution time is obtained. For a
two-dimensional target system, we show how by appropriately combining
quantum feedback with dynamical decoupling methods, concatenated
feedback-decoupling schemes may be built, which can operate under
relaxed control assumptions and can outperform purely closed-loop and
open-loop protocols.
\end{abstract}


\maketitle

\section{Introduction}

The role of real-time measurement and information feedback in
designing control strategies is well recognized for a wide range of
applications in classical control theory \cite{doyle-feedback}.
Feedback in the quantum domain, however, unavoidably involves
additional challenges due to the intrinsically probabilistic
\cite{sakurai,holevo} and, typically, destructive character of the
quantum measurement process \cite{vonneumann} -- leading to a host of
delicate issues and added difficulties for many tasks of interest to
the emerging field of quantum control. As a result, real-time quantum
feedback is nowadays actively investigated at both the theoretical and
experimental level. In particular, continuous-time feedback
strategies, based on indirect weak measurements, have been originally
proposed for cavity QED systems
\cite{wiseman-milburn,wiseman-feedback}, and have recently begun to be
experimentally accessible \cite{mabuchi-science,orozco}.  From a
control-theoretic standpoint, a general analysis of strategies based
on average measurements on quantum ensembles and feedback has been
considered in \cite{altafini-feedback}.  Suggestive applications of
continuous measurements and feedback in the context of quantum error
correction have also been investigated \cite{ahn-feedback}. 

Our focus in this work is to begin exploring possible uses and
advantages of discrete-time quantum feedback, based on strong
projective measurements, in connection with a fundamental quantum
control task: engineering a desired unitary operation on a target
system, which may be generally interacting with an uncontrollable
quantum environment.  Thus, given a set of available control
operations, the goal is to design a protocol capable to enforce a
desired net evolution on the system at a given final time, {\em
irrespective of the system's initial state}. This problem, unlike the
simpler case of {state preparation}, forces in general the
introduction of {\em indirect measurements} on a suitable ancillary
system.  The first step in this direction is to employ the minimum
physical amount of information we can gather, {\em a single bit}. In
particular, in what follows, we aim to construct and test protocols
for engineering the simplest unitary operation (the identity or {\tt
no-op} gate \cite{nielsen-chuang}) by using, along or in conjunction
with additional control schemes, a feedback loop built from basic,
arbitrarily fast and accurate control operations.

The analysis and techniques presented here exploit and extend ideas
close to those introduced in a number of earlier works. In particular,
our ancillary qubit system may be seen as a concrete instance of a
quantum controller \cite{lloyd-coherentqfeedback} or, more generally,
a universal quantum interface \cite{viola-engineering,lloyd-uqi}.
Control schemes exploiting auxiliary quantum probe systems and
measurement capabilities have also been recently discussed in
\cite{katerina,domenico}.  However, while the main emphasis of the
above-mentioned literature is on establishing general existential
results on quantum accessibility and controllability, our interest
here is on providing constructive, synthesis results instead.

Two main families of protocols emerge: on one hand, {\em
feedback-enacted decoupling protocols} -- which, in the infinitesimal
time limit, directly relate to feedback implementations of quantum
dynamical symmetrization procedures
\cite{viola-1,zanardi-symmetrizing} and, when appropriate conditions
are met, allow to relax the fast control limit of open-loop schemes;
on the other hand, {\em feedback-concatenated decoupling protocols} --
whereby the output of a primary open-loop dynamical decoupling (DD)
protocol serves as input to an outer feedback block, or viceversa.
Interestingly, the analysis of the feedback-enacted procedures may be
also regarded as stemming from the communication-oriented approach of
\cite{gregoratti-werner,gregoratti-werner-2}, or as a particular
instance of a quantum error-correction procedure \cite{howardQEC}. In
a similar spirit, an interesting discussion on the ensuing trade-off
between information gain and disturbance for different measurement
strategies in a single-qubit setting can be found in
\cite{doherty-singlequbit}.  From a technical point of view, our main
results further demonstrate the usefulness of {\em linear algebraic
methods} for finite-dimensional quantum control settings, along the
lines introduced in \cite{ticozzi-lineareqDD}.

The content of the paper is organized as follows.  Following a
presentation of the general control-theoretic setting in Section
\ref{setting}, we develop our main theoretical analysis and results
for a generic finite-dimensional target system in Section
\ref{theory}. By restricting to the simple yet paradigmatic case of
arbitrary state stabilization for a single qubit, in Section
\ref{application} we describe various concatenated feedback-decoupling
protocols built, in particular, from both maximal and selective DD
schemes \cite{viola-1,highresNMR}.  The performance of different
schemes in the non-ideal limit of finite control time scales is
numerically investigated and compared in Section \ref{numerical}.  A
discussion and concluding remarks follow.

\section{System and control settings}\label{setting}

We will consider a finite-dimensional quantum system $\Si$, coupled in
general to an uncontrollable, possibly unknown quantum environment
$\Ei$. Let $\Hi_\Si$ and $\Hi_\Ei$ denote the system and the
environment Hilbert spaces, with dim($\Hi_\Ai$)= $d_\Si$,
dim($\Hi_\Ei$)= $d_\Ei$, respectively. Assume that the joint system
plus environment dynamics is driven by a time-independent Hamiltonian
of the form: \bea\label{hamiltonian} H_{tot}=H_\Si\otimes
I_\Ei+I_\Si\otimes H_\Ei+H_{\Si\Bi}
\sum_{i=0}^k
S_i\otimes B_i\,, \eea
\noindent
where $k$ is a finite integer, $H_\Si,H_\Bi,H_{\Si\Ei}$ are the
system, environment, and interaction Hamiltonians, and $S_i$, $B_i$
are arbitrary Hermitian operators on $\Hi_\Si$ and $\Hi_\Ei$,
respectively. Without loss of generality, we may identify $S_0=
I_{\cal S}$ and take the $S_i$ to be traceless for $i=1,\ldots,k$.

We will assume access to the following control resources and
capabilities:
\begin{itemize}
\item An additional two-level system $\A$ with Hilbert space $\Hi_\Ai$
(an \emph{ancilla system}, or a \vv{quantum information channel}
henceforth), able to be prepared in a given pure state
$|\phi\rangle$. We will assume $\A$ to have a decoherence time much
longer than the relevant control time scale, so that it can be treated
as a closed system, evolving under the trivial Hamiltonian $H_{\cal
A}=0$ under ideal conditions.

\item Arbitrarily {\em fast and strong unitary transformations} on the
system and the ancilla;

\item {\em Fast conditional gates}, i.e. unitary operations of the
form $C_{U_\Si}^\Ai=\ket{0}\bra{0}_\Ai\otimes
I_\Si+\ket{1}\bra{1}_\Ai\otimes U_\Si$ on $\Hi_\Ai \otimes \Hi_\Si$ --
$\{\ket{0},\ket{1}\}$ denoting an orthonormal basis in $\Hi_\Ai$.

\item {\em Projective \vv{Yes-No} measurements} on the ancilla,
corresponding to the projections on the reference basis
subspaces. Suppose, moreover, that we can read, record, and use the
\emph{classical information} we obtain to influence the subsequent
control strategy.

\end{itemize}

In the reminder of this section, we recall some basic results related
to DD and symmetrization, referring the reader to
\cite{viola-1,zanardi-symmetrizing,viola-engineering,viola02} for more
detailed discussions.

DD strategies for open quantum systems are based on the idea of
coherently modify the target dynamics by interspersing the natural
evolution with (ideally) instantaneous control manipulations acting on
the system alone. Let $H_c(t)\otimes I_\Ei$ denote the applied,
semiclassical control Hamiltonian.  DD is most conveniently described
by effecting the time-dependent change of basis in $\Hi_\Si$ induced
by the control propagator $U_c(t)={\cal T}\exp\{-i \int^t_0 H_c(s)
ds\}$ (transforming to the so-called ``logical frame'', and setting
$\hbar =1$ henceforth).  For deterministic DD, control sequences are
additionally assumed to be periodic, meaning that $U_c(t+T_c)=U_c(t)$
for a cycle time $T_c>0$.  Then, the total transformed Hamiltonian
reads:
\begin{equation}\tilde{H}(t)=\sum_{i=0}^k
U^\dag_c(t)S_iU_c(t)\otimes B_i\,.
\end{equation} 
The associated logical propagator $\tilde{U}(t)$, which coincides with
the full Schr\"odinger propagator at every instant $T_N= NT_c$, $N \in
{\mathbb N}$, may be conveniently expanded in the Magnus series
\cite{magnus},
\begin{equation}\label{mag}
\tilde{U}(T_N)= \left[ e^{-i {\bar H}T_c }\right]^N\,, \;
\bar H = \bar H^{(0)}+\bar H^{(1)}+\bar H^{(2)}+...\,,
\end{equation}
where ${\bar H}$ denotes the average Hamiltonian \cite{highresNMR}.
Explicitly, the lowest-order terms are given by
\beq\label{Hfirst}
\bar{H}^{(0)}=\frac{1}{T_c}\int_0^{T_c}ds \tilde{H}(s)\,,\eeq
\noindent
\beq\label{Hsecond}   \bar{H}^{(1)}=-\frac{i}{2T_c}\int_0^{T_c}   ds_1
\int_0^{s_1} ds_2\, [\tilde{H}(s_2), \tilde{H}(s_1)]\,, \eeq
\noindent
with higher-order terms expressible as a power series in the control
parameter $T_c$.  Given a finite evolution time $T>0$, DD is
implemented by considering a time-slicing into $N$ control cycles over
$[0,T]$, with $T\equiv T_N=NT_c$.  In the ideal limit of {\em
arbitrarily fast control}, $N\rightarrow\infty$, $T_c \rightarrow 0$,
it can be shown the higher order terms in (\ref{mag}) are negligible
in a suitable norm, and the dominant contribution to the logical
propagator takes the simple form
$$\tilde{U}(T_N)\approx e^{-i\bar H^{(0)}T_N}\,.$$ 
\noindent
In practice, higher-order terms turn out to be critical in affecting
the control performance when the ideal condition $N\rightarrow\infty$
is relaxed.  The ideal limit still proves, however, a useful starting
point. Suppose we choose a piecewise-constant control propagator,
$$U_c(t)\equiv G_j\,, \;\; j\Delta t\leq t\leq (j+1)\Delta t\,,$$ 
\noindent
with $j=0,\ldots,n_g-1,$ $1 < n_g \in {\mathbb N}$, and $\Delta
t=T_c/n_g$.  From Eq.~\eqref{Hfirst}, this control prescription gives
the following overall average Hamiltonian:
\begin{eqnarray}\label{DD}
\bar{H}^{(0)} &=& \sum_i\bigg(\frac{1}{n_g}\sum_{j=1}^{n_g}G_j^\dag
S_i G_j\bigg)\otimes B_i \\ & = & I_{\cal S}\otimes H_{\cal E} +
\sum_{i >0} \bigg(\frac{1}{n_g}\sum_{j=1}^{n_g}G_j^\dag S_i
G_j\bigg)\otimes B_i \,.\nonumber\end{eqnarray} 
\noindent
By ensuring that the cyclicity requirement is fulfilled (implying that
$\prod_j G_j$ $=I_{\cal S}$), the above Hamiltonian specifies the
lowest-order stroboscopic evolution in the physical frame also.  In
practice, for a large class of DD schemes the control operations $G_j$
form a discrete group ${\cal G}$, with $n_g=|{\cal G}|$.  In this
case, it is possible to show that the quantum operation in parenthesis
in Eq.~(\ref{DD}) effectively projects onto system Hamiltonians which
are invariant under ${\cal G}$ -- hence commute with the $G_j$
themselves.

In what follows, we will exploit DD to synthesize a vanishing average
Hamiltonian, i.e. to \vv{freeze} the dynamics to first- (or higher)
order in time. Let us recall two relevant DD schemes for a single
two-level system, which will be useful in Section \ref{application}.

$(i)$ First, suppose the control objective is to suppress evolution
due to a known drift and/or a known coupling Hamiltonian which,
without loss of generality, may be assumed to be along $z$. Thus, the
non-trivial $S_i$ in (\ref{hamiltonian}) are proportional to the Pauli
matrix $\sigma_z$.  Then, DD over $[0,T_c]$ is achieved by letting
${\cal S}$ evolve till $T_c/2$, then applying an instantaneous
transformation $\sigma$ such that
$\sigma\sigma_z\sigma^\dag=-\sigma_z$ (e.g., $\sigma = \sigma_x$), and
applying a second identical transformation after another $T_c/2$ to
complete the cycle.  The desired averaging is ensured by the property
$\sigma_z +\sigma\sigma_z\sigma^\dag=0$.  To improve performance for
finite $T_c$, a symmetrized version of the above sequence (so-called
Carr-Purcell DD) ensures that {\em odd} corrections in the Magnus
series are automatically eliminated. This is attained with a slight
re-arrangement of the control time intervals: Let ${\cal S}$ evolve
freely till $t=T_c/4,$ set the control propagator to $\sigma$ and let
it constant until $t=3T_c/4$; then reset the control propagator to
$I_{\cal S}$ and let ${\cal S}$ evolve for the remainder of the cycle.

$(ii)$ If no information is available about the drift and/or coupling
Hamiltonian to be suppressed, a {\em maximal} DD strategy need to be
employed. As shown in \cite{viola-1}, cycling the control propagator
through the elements of an orthonormal basis for $\mathfrak{su}(2)$,
i.e. $I,\sigma_x,\sigma_y,\sigma_z$, ensures that the resulting
average Hamiltonian is always zero, reflecting the fact that the
action of the Pauli group is irreducible.  Interestingly, while the
order in which the control propagators are switched along the cycle is
irrelevant to lowest-order averaging, different ``control paths'' over
the same set may lead to different control performance for finite
$T_c$, due to their different influence on higher-order corrections.

\vfill


\section{Feedback-Enacted Dynamical Decoupling}\label{theory}

Our first goal is to analyze the possible advantages of using a
classical, single-bit feedback strategy to engineer the identity
evolution on $\Si$. It is apparent that projective measurements on
${\cal S}$ itself cannot be of any use toward realizing a unitary
operation, as in general the pre-measurement state is irreversibly
lost after the information is gathered. Thus, as anticipated, we will
consider a combination of idealized, instantaneous unitary control
actions on the pair ${\cal S}$ plus ${\cal A}$, and projective
measurements on ${\cal A}$, to dynamically eliminate unwanted
components.

Assume that ${\cal A}$ is prepared in an initial state
$\ket{\phi}=a\ket{0}+b\ket{1}$, with coefficients $a,b$ to be
determined. Without loss of generality, we may also assume that the
${\cal E}$ is initially in a pure, though unknown in general, state --
denoted by $\ket{\xi}$. Finally, let $|\psi\rangle$ denote an arbitrary
pure state of ${\cal S}$. The basic steps of a {\em Feedback-Enacted
DD protocol} ({\tt FDD}) may be described as follows:

(I) Rapidly entangle ${\cal S}$ and ${\cal A}$, by performing a
conditional gate $U_\Si$: \bea C_{U_\Si}^\Ai\ket{\phi}\ket{\psi}
=a\ket{0}\ket{\psi}+b\ket{1}U_\Si\ket{\psi}.\eea

(II) Evolve freely, in the presence of ${\cal E}$, up to a finite time
$T$. The combined final state may be described as follows:
\bea\label{evolution} I_\Ai\otimes
U_{\Si\Bi}\left(a\ket{0}\ket{\psi}\ket{\xi}+b\ket{1}
U_\Si\ket{\psi}\ket{\xi}\right)=\nonumber\\
=a\ket{0}U_{\Si\Bi}\ket{\psi}\ket{\xi}+b\ket{1}
U_{\Si\Bi}\left(U_\Si\ket{\psi}\right)\ket{\xi}\,.  \eea

(III) Proceed by undoing the preparation step, that is, effect the
following conditional transformation: \bea C_{U^\dag_\Si}^\Ai\otimes
I_\Ei(a\ket{0}U_{\Si\Bi}\ket{\psi}
\ket{\xi}+b\ket{1}U_{\Si\Bi}(U_\Si\ket{\psi})\ket{\xi})=\nonumber\\
a\ket{0}U_{\Si\Bi}\ket{\psi}\ket{\xi}+b\ket{1}(U^\dag_\Si\otimes
I_\Ei)U_{\Si\Bi}(U_\Si\otimes I_\Ei)\ket{\psi}\ket{\xi}. \eea
\noindent 
The above equation makes it clear that, in order to obtain a {\em
balanced} averaging of the two unitary components, equal weights on
the two ancilla states are needed.  We thus set $a=b=1/\sqrt{2}$
henceforth.

(IV) Apply a Hadamard transformation $U_H^\Ai$ \cite{nielsen-chuang}
on ${\cal A}$, obtaining the combined state
\begin{widetext}
\bea \label{premeas}&&\frac{1}{\sqrt{2}}U_H^\Ai\otimes
I_{\Si\Ei}\left[\ket{0}U_{\Si\Bi}\ket{\psi}\ket{\xi}+\ket{1}(U^\dag_\Si\otimes
I_\Ei)U_{\Si\Bi}(U_\Si\otimes I_\Ei)\ket{\psi}\ket{\xi}\right]=\nonumber\\
&&=\frac{1}{{2}}\ket{0}\left(U_{\Si\Bi}+(U^\dag_\Si\otimes
I_\Ei)U_{\Si\Bi}(U_\Si\otimes I_\Ei)\right)\ket{\psi}\ket{\xi}+
\frac{1}{{2}}\ket{1}\left(U_{\Si\Bi}-(U^\dag_\Si\otimes
I_\Ei)U_{\Si\Bi}(U_\Si\otimes I_\Ei)\right)\ket{\psi}\ket{\xi}.\eea
\end{widetext}

(V) Measure the ancilla, thereby selecting the effective evolution on
the joint ${\cal H}_\Si\otimes{\cal H}_\Ei$ space.  Such a conditional
evolution will not be unitary in general.  In particular, if the
outcome is $0$, then the conditional joint state may be seen as the
result of the following generalized measurement operation:
$$\rho_{\Si\Bi}(T) \Big|_0=\frac{1}{ p_0(T)}\left(A_{\Si\Bi}^{(+)}(T)
\rho_{\Si\Bi}(0) A^{(+)\dagger}_{\Si\Bi}(T) \right)\,,$$
\noindent 
where
$\rho_{\Si\Bi}(0)=\ket{\psi}\bra{\psi}\otimes\ket{\xi}\bra{\xi},$ and
\beq A^{(\pm)}_{\Si\Bi}=\frac{1}{2}\left( U_{\Si\Bi}\pm
(U^\dag_\Si\otimes I_\Ei) U_{\Si\Bi}(U_\Si\otimes I_\Ei)\right)\,,
\end{equation}
\beq
\label{prob0}
p_0(T)\equiv p(T)\Big|_0=\Tr\left(A^{(+)\dagger}_{\Si\Bi}(T)
A^{(+)}_{\Si\Bi} (T)\rho_{\Si\Bi}(0) \right),\eeq
is the corresponding conditional evolution probability at time $T$.
From Eq.~\eqref{premeas}, we notice that the net effect of the above
(I)--(V) protocol is similar to a \emph{two-steps} DD strategy
\cite{viola-1,ticozzi-lineareqDD}, where, however, the role of the
unwanted coupling Hamiltonian is taken by the joint ${\cal SE}$
propagator.

From a control standpoint, one is naturally led to the following
design issue: Given a joint propagator $U_{\Si\Ei}(T)$ for arbitrary
$T>0$, find unitary matrices $U_\Si$ and $U_{\tt fb}$ on the system,
such that the {\em conditional decoupling conditions} are satisfied,
\begin{eqnarray}
&&
A^{(+)}_{\cal SE}(T) =I_{\cal S}\otimes B_{\cal E}^{(+)}(T)
\,,\label{UDD} \\ && A^{(-)}_{\cal SE}(T)=U^\dag_{\tt fb}\otimes
B_{\cal E}^{(-)}(T)\,,\label{UDD1}
\end{eqnarray} 
where $B_{\cal E}^{(\pm)}(T)$ account for the effect of the dynamics
on ${\cal E}$, and $U_{\tt fb}$ conditional to the outcome
$\ket{1}_{\cal A}$ corrects the unintended component.
Notice that if condition~\eqref{UDD} can be satisfied with $B_{\cal
E}^{(+)}$ unitary, the probability of obtaining $\ket{0}_{\cal A}$ is
automatically one.  In practice, although {\em no} constraint on the
evolution time exists, ensuring perfect knowledge of the joint
propagator $U_{\Si\Bi}(T)$ may be hard to achieve.  We are thus
brought to determine conditions on $U_{\Si\Ei}(T)$ that may still
allow a solution of the above DD problem.

\subsection{Short time scale: Necessary and sufficient conditions}

Assume first that the relevant evolution time is sufficiently short
(formally infinitesimal), so that a first-order Taylor expansion
accurately represents the joint propagator $U_{\Si\Ei}$, \beq
U_{\Si\Ei}\approx I_{\cal SE}-i H_{\Si\Ei}\delta t +{\cal O}(\delta
t^2)\,.
\label{firstorder}
\eeq 
\noindent
The \vv{decoupling-like} conditions we are seeking may be derived by
substituting \eqref{firstorder} in \eqref{UDD}.  Due to the
tracelessness assumption on $S_i, i>0$, the latter equation is
equivalent to
$$ \sum_{i>0} \left(S_i + U^\dag_{\cal S} S_i U_{\cal S}\right)
\otimes B_i = 0\,.$$ 
The following results, which extends \cite{ticozzi-lineareqDD}, 
is relevant.

\begin{thm}\hspace{-1mm}{\bf.} \label{mixingthm} Let $X$ be 
an $n$-dimensional, traceless normal matrix and {\em
$\spec(X)=\{x_i\}_{i=1}^n$} its spectrum. Then, the equation
\beq\label{eq1} X+U^\dag XU =0\,, \eeq admits a unitary solution $U$
if and only if there exists a permutation of the elements of
{\em$\spec(X)$} such that the following {\text mixing condition} is
satisfied:
\begin{equation}\label{mixing}
x_i=-x_{n-i+1},\quad i=1,\ldots,n.
\end{equation}
If, in addition, $X$ is Hermitian, and the spectrum elements are taken
with descent ordering, then condition \eqref{mixing} implies a
solution of the form \beq\label{sol} U=V JV^\dag\,, \eeq where $V$ is
a unitary matrix satisfying $V^\dag XV=\tilde{X}\equiv{\rm
diag}(x_1,\ldots,x_n)$, and
$$J=
\left[
\begin{array}{cccc}
 0 &  0 & \dots &1  \\
 0 &  \dots & 1&0   \\
 0 & \begin{array}{ccc}&&\!\!\!\!\!\! .\\[-3mm]
 &\!\!\!\! .&\\[-3mm]
 .&&\end{array}  &
 \begin{array}{ccc}&&\!\!\!\!\!\! .\\[-3mm]
 &\!\!\!\! .&\\[-3mm]
 .&&\end{array} &\vdots \\
1 & \ 0& \dots &0
\end{array}
\right].
$$
\end{thm}

\proof First, notice that Eq. (\ref{eq1}) has a unitary solution $U$
if and only if \beq\label{eq2} \tilde{X}+\tilde{U}^\dag
\tilde{X}\tilde{U} =0 \eeq has a unitary solution $\tilde{U}$ (in
which case, $\tilde{U}=V^\dag UV$).  For the {\em if} implication,
assume that \eqref{mixing} is satisfied for a permutation of the first
$m$ integers $(i_1,\ldots,i_m)$.  Then the unitary matrix $\tilde{U}$
which reorders the basis vectors indexes according to that
permutation, i.e.,
$$\tilde{U}\ket{\psi_j}=\ket{\psi_{i_j}},\quad j=1,\ldots,m,$$
satisfies \eqref{eq2}.  For proving the opposite implication, assume
that $\tilde{U}$ is a solution of (\ref{eq2}). Then $\tilde{U}^\dag
\tilde{X}\tilde{U}$ must be diagonal. Moreover, since a unitary change
of basis does not change the spectrum, $\tilde{U}^\dag
\tilde{X}\tilde{U}={\rm diag}(x_{i_1},\ldots,x_{i_m})$, where
$(i_1,\ldots,i_m)$ is a permutation of the first $m$
integers. Employing again (\ref{eq2}), we get $\tilde{X}=-{\rm
diag}(x_{i_1},\ldots,x_{i_n}),$ hence condition (\ref{mixing})
holds. The specialization to the Hermitian case is
straightforward. \cvd

The above theorem fully characterizes the interactions able to be
suppressed via a two-steps DD scheme: In fact, all the operators $S_i,
i>0$ need to satisfy the condition of the theorem, in the {\em same
basis}, that is, employing the same unitary change of basis $V$.
Moreover, the form of the control actions to be employed is
constructively provided in the proof.

Assume that the relevant conditions are satisfied.  Then we can choose
$U_\Si$ such that
\begin{eqnarray}
&&A^{(+)}_{\cal SE}(\delta t)=I_{\cal S}\otimes (I_{\cal E}-i H_{\cal
E}\delta t)\,, \nonumber \\ &&A^{(-)}_{\cal SE}(\delta t)=i
\Big(\sum_{i>0} S_i \otimes B_i\Big) \, \delta t\,.
\end{eqnarray}

\noindent Thus, in this limit, \beq \lim_{\delta t\rightarrow
0}p(\delta t)\Big|_1 = {\cal O}(\delta t^2) \rightarrow 0\,,
\eeq 
\noindent 
meaning that the faulty outcome $1$ has zero probability to be
obtained -- consistent with the fact that $B^{(+)}_{\cal E}(\delta t)$
is unitary.

In principle, evolution over a finite time $T$ may always be obtained
by iterating the the basic steps (I)--(V) a sufficiently large number
of times $N$, so that $\delta t =T/N \rightarrow 0$.  The joint
probability of obtaining zero as result of {\em all} the measurements
up to $T$ becomes then
\begin{eqnarray} p(T)\Big|_{0,\ldots,0}&=&\lim_{N\rightarrow
\infty}\Big\|\left(A^{(+)}_{\Si\Bi} (\delta t)\right)^N\ket{\psi}
\ket{\xi}\nonumber\Big\|^2\\ & \approx &
\lim_{N\rightarrow\infty}\Big\|\Big(I-\frac{iT}{N} H_{\cal
E}\Big)^N\ket{\psi}\ket{\xi}\Big\|^2\nonumber\\
&\approx&\left\|e^{-i H_{\Bi} T}
\ket{\psi}\ket{\xi}\right\|^2\nonumber = 1\,,\end{eqnarray} 
\noindent 
as expected.  In this form, our {\tt FDD} protocol may be thought of
as achieving an implementation of DD directly based on the quantum
Zeno effect \cite{facchi-zeno}.  Formally, the protocol may also be
regarded as a special case of the dynamical symmetrization scheme via
repeated measurements originally proposed by Zanardi
\cite{zanardi-symmetrizing}.  While we have intentionally restricted
the analysis to a two-dimensional ancilla (hence group-algebra in the
language of \cite{zanardi-symmetrizing}), this allows an {\em explicit
characterization} (via Theorem 1) of the open-system Hamiltonians for
which the protocol is effective.  In addition, as it will be clear in
the next Section, one of our main goals is to explore the use of
classical feedback blocks to enhance the performance of open-loop DD.

Before moving to consider finite-time feedback evolutions, we address
the robustness of the {\tt FDD} protocol against unintended
Hamiltonian drifts on the ancilla subsystem, $H_{\cal A}\not =0$. By
starting from the same initial state for ${\cal SA}$ as in step (I)
above, in addition to the joint system-environment evolution
$U_{\Si\Ei}$, we now also consider a unitary evolution $U_\Ai$ on
${\cal A}$. It suffices to describe its action on the basis states:
\begin{eqnarray}
\ket{0}&\mapsto&
\cos{\theta}\ket{0}+\sin{\theta}e^{i\phi}\ket{1}\nonumber\\
\ket{1}&\mapsto& -\sin{\theta}\ket{0}+\cos{\theta}e^{-i\phi}\ket{1}\,,
\end{eqnarray}
where $\theta,\phi$ are linear functions of $t$.  Thus, the joint
state after step (II) is
\begin{eqnarray*}
\ket{\psi'}&=&(\cos{\theta}\ket{0}+\sin{\theta}e^{i\phi}\ket{1})
U_{\Si\Ei}\ket{\psi}\ket{\xi}\\
&+&(-\sin{\theta}\ket{0}+\cos{\theta}e^{-i\phi}\ket{1})
U_{\Si\Ei}(U_\Si\ket{\psi})\ket{\xi}.
\end{eqnarray*}
Applying $C^\Ai_{U^\dag}$ and $U^\Ai_H$ as in steps (III)-(IV),
respectively, yields:
\begin{widetext}
\begin{eqnarray*}\ket{\psi''}&=&\ket{0}\left(\cos{\theta}U_{\Si\Ei}+
\sin{\theta}e^{i\phi}(U_\Si^\dag\otimes I_\Ei)U_{\Si\Ei}-\sin\theta 
U_{\Si\Ei}(U_\Si\otimes
I_\Ei)+\cos\theta e^{-i\phi}(U_\Si^\dag\otimes
I_\Ei)U_{\Si\Ei}(U_\Si\otimes I_\Ei)\right)\ket{\psi}\ket{\xi}\\
&+&\ket{1}\left(\cos{\theta}U_{\Si\Ei}-\sin{\theta}e^{i\phi}(U_\Si^\dag\otimes
I_\Ei)U_{\Si\Ei}-\sin\theta U_{\Si\Ei}(U_\Si\otimes
I_\Ei)-\cos\theta(U_\Si^\dag\otimes I_\Ei)U_{\Si\Ei}(U_\Si\otimes
I_\Ei)\right)\ket{\psi}\ket{\xi}.\end{eqnarray*}
\end{widetext}
Finally, if the joint probability for obtaining always $\ket{0}$ as a
result of the measurements is computed, to first order in $t$ one
obtains
\begin{eqnarray*} 
&& \hspace*{-2mm}p(T)\Big|_{0,\ldots,0} \approx \lim_{N\rightarrow
\infty}\Big\|\left(A^{(+)}_{\Si\Bi}(\delta t; \theta, \phi)\right)^N\ket{\psi}
\ket{\xi}\nonumber\Big\|^2 \approx\nonumber \\
&&
\hspace*{-2mm}\lim_{N\rightarrow\infty}\Big\|\Big[I-\frac{iT}{N}\Big(
H_{\cal E}+ \frac{\phi_0}{2} + \frac{(U_{\cal S}^\dag -U_{\cal S})}{2i} 
\theta_0 
\Big)
\Big]^N\ket{\psi}\ket{\xi}\Big\|^2\,,\nonumber\\\end{eqnarray*} where
$\phi_0=\dot{\phi}(0),\theta_0=\dot{\theta}(0)$, respectively.
Accordingly, if $U_\Si$ is both unitary and Hermitian (as it is the
case e.g., for a CNOT gate), the same expression of the ideal case is
recovered, and the above probability still tends to $1$.  Since, under
these conditions, the unitary drift of the ancilla is irrelevant, this
relax the requirement of a \vv{frozen} ancilla for sufficiently fast
protocols.

\subsection{Finite evolution time: Necessary and sufficient conditions}

We now provide necessary and sufficient conditions on the joint ${\cal
SE}$ propagator for employing the {\tt FDD} protocol {\em one-shot}
over a finite time interval.  We begin with a definition.

\begin{defin}\hspace{-1mm}{\bf .}\label{LD}
An operator $X_{\Si\Bi}$ acting on a bipartite Hilbert space
$\Hi_{\Si\Bi}=\Hi_\Si\otimes\Hi_\Bi$ is {\em locally diagonalizable}
(LD) if there exist operators $\{A_j\}_{j}$ and $\{B_j\}_{j}$ on
$\Hi_\Si$ and $\Hi_\Bi$, respectively, such that
\begin{eqnarray}
&&X_{\Si\Bi}=\sum_j A_j\otimes B_j\,,\;\, with \nonumber \\
&&[A_i,A_j]=0,\quad \forall i,j\,.\label{locdiag}
\end{eqnarray}
\end{defin}

The fact that an operator on a tensor product space can always be
decomposed in the form $X_{\Si\Bi}=\sum_j A_j\otimes B_j$ follows for
example from the operator-Schmidt decomposition
\cite{nielsen-operatorschmidt}. Condition \eqref{locdiag} ensures that
there exist a unitary change of basis, corresponding to an operator
$\hat U_\Si$ in $\Hi_\Si$, that simultaneously diagonalizes all the
$A_j$'s, that is, $\hat U_\Si$ brings $U_{\Si\Bi}$ to a block-diagonal
form.

Clearly, all factorizable normal operators on finite-dimensional
Hilbert space are LD.  However, the converse is not true.  For
Hamiltonians as in Eq.~\eqref{hamiltonian}, all evolutions generated
under the condition that $[H_{\cal S}, H_{\cal SE}]=0$ are LD.  This
includes, for instance, purely dephasing spin-bath or spin-boson
models as in \cite{viola-seminalDD}.  In the language of quantum error
correction, LD open-system evolutions correspond to purely Abelian
error algebras \cite{viola-generalnoise}.
  
\begin{prop}\hspace*{-1mm}{\bf .}
\label{qubitfeedback}
Let $\textrm{\emph{dim}}(\Hi_\Si)=2$, and let the joint evolution with
the environment $\Bi$ at time $T$ be described by the propagator
$U_{\Si\Bi}(T)$. There exists a one-bit feedback DD protocol that
restores the initial state of the system for arbitrary $T$ if and only
if $U_{\Si\Bi}$ is LD.
\end{prop}

\proof For convenience, assume now the ordering $\Bi\otimes\Si$ in the
tensor products. By hypothesis, $U_{\Si\Bi}$ satisfies
\eqref{locdiag}, thus we may work in a basis where all the $A_j$ are
diagonal by applying a suitable $\hat U_\Si$.  In such a basis each
block of $U'_{\Si\Bi}=\hat U_\Si U_{\Si\Bi} \hat U_\Si^\dag$ becomes
diagonal. Consider then these diagonal blocks, that is, the
${d_\Bi}^2$ $2\times 2$-matrices
$$ \left(%
\begin{array}{cc}
  d_{1,i} & 0 \\
  0 & d_{2,i} \\
\end{array}%
\right),\quad i=1,\ldots,{d_\Bi}^2\,.$$
Each of them may be rewritten as:
$$ \left(%
\begin{array}{cc}
  d_{1,i} & 0 \\
  0 & d_{2,i} \\
\end{array}%
\right)=\frac{d_{1,i}+d_{2,i}}{2}I_2+\frac{d_{1,i}-d_{2,i}}{2}\sigma_z.$$
Hence, the following decomposition applies:
$$ U'_{\Si\Bi}=B'_1\otimes I_2+B'_2\otimes\sigma_z,$$ where $B'_1,
B'_2$ are operators on $\Hi_\Bi$. It is now apparent from
Eq.~\eqref{UDD} that, for example, $U_\Si=\hat U_\Si\sigma_x\hat
U_\Si^\dag$ can {decouple} the system if the ancilla is found in the
$\ket{0}$ state, $\hat U_\Si$ being the change of basis that
diagonalizes the $A_j$. If the outcome $\ket{1}$ is measured, the
resulting evolution operator for the system and the bath in the
diagonal basis is of the form:
$$2 B'_2\otimes\sigma_z,$$ that is, factorized and unitarily
correctable in the original basis for ${\cal S}$ by employing $U_{\tt
fb}=\hat U_\Si\sigma_z\hat U_\Si^\dag$. Necessity of the LD condition
follows from the fact that each block must be proportional to the
identity simultaneously and from invoking Theorem
\ref{mixingthm}. \cvd

For dimension, $d_{\cal S}>2$, the algebraic conditions that need to
be satisfied become more demanding: Essentially, the same kind of
mixing structure found in Theorem \ref{mixingthm} has to hold for
$U_{\Si\Ei}$. In fact, the above Proposition may be seen as a
specialization of the following result:

\begin{thm}\label{thm2}\hspace{-1mm}{\bf .}
There exists a one-bit feedback-DD protocol that restores the initial
state of the system for arbitrary $T$ if and only if $U_{\Si\Bi}$ is
LD with each $d_{ \cal S}$-dimensional block of the form:
\beq\label{blocks} 
\left(
\begin{array}{cccc}
  d_{1,i} & 0 & \cdots & 0 \\
  0 &  d_{2,i} & 0 & \vdots \\
  \vdots & 0 & \ddots & 0 \\
  0 & \cdots & 0 & d_{d_{\cal S},i} \\
\end{array}%
\right)= \left(\frac{1}{d_{\cal S}}\sum_{j=1}^{d_\Si}
d_{j,i}\right)I_{d_{\cal S}}+c_i\bar D\,,\eeq
\noindent
where $c_i$ is a coefficient depending on the block index, and $\bar
D$ is unitary and satisfies the spectral conditions of Theorem 1.
\end{thm}

\proof Since $U_{\Si\Bi}$ is LD and Eq.~\eqref{blocks} holds, the fact
that $\bar D$ satisfies \eqref{mixing} is a sufficient and necessary
condition for the existence of an effective $U_\Si$ by use of Theorem
\ref{mixingthm}. Here, however, we deal with complex eigenvalues, so
there is no descent ordering and we cannot give the form of decoupling
operation $U_\Si$ in a natural way. It has now to be proven that there
exists a unitary operation working for the feedback correction step if
the {faulty} outcome has been measured. From Eq.~\eqref{UDD1}, this is
true if and only if $\bar D$ is proportional to a unitary operation,
i.e. all its eigenvalues have the same absolute value. Including the
coefficient in $c_i$ yields the desired result.  \cvd


The {\tt FDD} strategy has a potentially important advantage with
respect to the standard open-loop averaging schemes, provided that the
appropriate algebraic conditions are fulfilled: Its effectiveness does
{\em not} directly depend on the elapsed interaction time $T$.  In
practice, of course, several characteristic time scales, depending on
both the open-system and control operations, will still be
relevant. In particular, we assumed to be able to enact ideal,
instantaneous unitary preparation, measurement, and resetting steps
throughout.  In practical terms, this will require the capability of
implement fast conditional gates and measurements with respect to the
typical correlation time scale of the underlying open-system dynamics.
The other strong assumption we made is to be able to access an
effectively decoherence-free ancilla \cite{footnote}.

\section{Feedback-Concatenated Dynamical Decoupling}
\label{application}

In the previous Section, we developed necessary and sufficient
linear-algebraic conditions for the unitary, joint evolution operator
for the target system plus environment $U_{\cal SE}$ to be fully
correctable by a feedback strategy involving a combination of fast
conditional operations and projective measurements.  However, making
sure that such conditions are verified, and explicitly constructing
the required control operations, demands in general an accurate
knowledge of both the form and the duration of the interaction --
restricting the usefulness of the procedure in realistic scenarios.
If such a detailed information is not available, our next goal is to
show how one can still make a feedback strategy effective by combining
the feedback action with open-loop averaging schemes -- in particular,
by appropriately modifying the evolution between subsequent
measurements.

In the rest of our present discussion, we will develop a general
solution in the simple yet paradigmatic case of a single qubit
($d_{\cal S}=2$) as the system of interest.  Possible extensions to
higher-dimensional systems will be mentioned in the conclusions.

\subsection{Problem setting}

For the sake of simplicity, we will describe in detail the basic
strategy for the Hamiltonian suppression case, that is, we set $B_i
=0$ for all $i$ in Eq.~\eqref{hamiltonian}.  Notice that this limit
could still account for non-unitary effects from a semiclassical
environment if randomly fluctuating parameters are allowed in the
Hamiltonian -- prominent physical examples being decoherence from
random telegraph noise \cite{Faoro} and from a dipolarly coupled
nuclear spin reservoir \cite{Merkulov}.  The proof for the general
open quantum system case relies on Proposition
\ref{qubitfeedback}. The control problem for which we will provide a
solution may then be stated as follows:
\begin{prob}\hspace*{-1mm}{\bf .}
\label{prob1} 
Consider a single qubit in the state $\rho(0)$ at $t=0$, undergoing an
{\rm uncertain} Hamiltonian evolution.  Assume we have an error set
$\mathcal{V}$ of possible Hamiltonians $\{H \}$, and an initial
estimate $\hat H \in\mathcal{V}$ for $H$. The task is to ensure that
$\rho(T)=\rho(0)$ at a given finite time $T$, for any possible
$H \in\mathcal{V}$.
\end{prob} 

This problem, which is equivalent to the implementation of a {\tt
no-op} unitary gate between times $0$ and $T$, qualifies as a {\em
robust} control problem.  That is, the output state must be
{insensitive} to errors in $\hat H$. The most general error
Hamiltonian may be described as an element of $i\cdot
\mathfrak{su}(2)$, that is, $\Delta H\equiv H-\hat H=
\vec{\epsilon}\cdot \vec{\sigma}=\epsilon_x\sigma_x+
\epsilon_y\sigma_y+\epsilon_z\sigma_z$. The initial estimate $\hat H$
is given by some \emph{a priori} knowledge, or guess, on the system's
behavior. It is often the case that approximated models are available
for the dynamics, and if such additional information is available, a
natural question is whether and how such information may be
efficiently exploited to improve over strategies which do not.  In our
case, a robust solution which invokes no ancilla and feedback loop,
and does not require in principle prior information, is provided by
the open-loop, maximal DD scheme described in Section II
\cite{viola-1}.  In fact, for both purely open-loop and feedback-based
strategies, the role of available {prior} information becomes critical
as soon as ideal control conditions are relaxed.  This will be
quantitatively addressed in Section \ref{numerical}.

\subsection{Concatenated control protocols}

Let as above $H=\hat H + \Delta H$ represent the target qubit
Hamiltonian in terms of an estimated plus error component.  Consider a
choice of axis and units such that $\hat H=\sigma_z$.  A first,
natural way to merge open- and closed-loop DD is to employ a selective
DD to remove the main component of the Hamiltonian, while relying on
feedback to counteract the residual ones. We will refer to the
resulting protocol as {\em Feedback-Enhanced Decoupling} ({\tt FED}).
From Section \ref{setting}, a single-qubit selective DD scheme based
on the control propagator sequence $\{I,\sigma_x\}$ projects onto the
$\sigma_x$ direction in the ideal limit of $T_c\rightarrow 0$.  Thus,
to first order, only an error component $\epsilon_x\sigma_x$ will be
left on the estimate.  The idea is to apply an ``outer'' block of {\tt
FDD}, so that a suitable choice of conditional operator, e.g.,
$C_{\sigma_z}^\Ai$ will correct for the above residual component.

In analogy to steps (I)--(V) in Section \ref{theory}, the {\tt FED}
protocol proceeds as follows: 

(I$'$) Prepare the ancilla in the initial state $|\phi\rangle
=(\ket{0}+\ket{1})/\sqrt{2}$ and entangle it with the qubit system by
applying the $C_{\sigma_z}^\Ai$ operation.

(II$'$) Evolve under a cycle of selective DD $\{I,\sigma_x\}$, by
obtaining at $T=T_c$ a net propagator of the form:
$$U_\Si^{res}(T_c)=e^{-i\phi\sigma_x}\,,$$ where $\phi=\epsilon_x
T_c$.

(III$'$)-(IV$'$) Undo the preparation steps applying the conditional
gate $C_{\sigma_z}^\Ai$ and a Hadamard $U_H^\Ai$, respectively.

(V$'$) Measure the ancilla in the $\{ |0\rangle, |1\rangle\}$ basis,
obtaining outcome ${0}$ with probability $p_0(T_c)=\cos^2(\phi)$, and
$1$ with probability $p_1(T_c)=\sin^2(\phi)$.

$U^{res}_{\cal S}(T_c)$ satisfies the conditions of Proposition
\ref{qubitfeedback}. In particular, it is easy to see that
$U_{\Si}=\sigma_z$ is a suitable operation for counteracting such a
residual evolution, irrespective of the value of $\phi$.  Hence, as
discussed previously, the classical information extracted from the
ancilla measurement restores the system's state in the \vv{0} case,
while in the other case it indicates univocally the correction needed
for taking the system back to the initial condition. In fact, applying
a fast resetting operation $\sigma_x^{\cal A}\otimes\sigma_x^{\cal S}$
when the measurement outcome is one, we restore the initial condition
of both ${\cal S}$ and ${\cal A}$ together. Remarkably, this happens
no matter how large is the error we are making in the prior, thereby
producing a maximal DD strategy in the ideal, fast control
approximation. This also means that the strategy is effective for
every Hamiltonian in $\mathcal{V}$, satisfying the robustness
requirement.

Note that at this level the role of the initial estimation does not
emerge clearly, suggesting another possible concatenation scheme in
place of the above {\tt FED} strategy.  The basic observation is that
nothing prevents us, in principle, from swapping the roles of the
feedback loop and the selective DD strategy.  Indeed, we may use a
$\{I,\sigma_z\}$ selective DD sequence to obtain a net evolution of
the form
$$U_\Si^{'\,res}(T_c)=e^{-i\phi'\sigma_z}\,,$$
\noindent
where now $\phi' = (1+\epsilon_z) T_c$, and employ $C_{\sigma_x}^\Ai$
operations along with a $\sigma_x^{\Ai}\otimes\sigma_z^{\Si}$
correction for the feedback loop. This choice leads to another
effective solution for our problem, in which we use the selective DD
to remove the errors, and the feedback loop to correct $\hat H$.  We
will refer to this second protocol as {\em Decoupling-Enhanced
Feedback} ({\tt DEF}). We will compare the effect of non-ideal
condition for both strategies with the aid of numerical simulations in
Section \ref{numerical}. A summary of the relevant strategies, along
with the respective roles of the feedback and open-loop components, is
given in Table \ref{table1}.

\begin{widetext}
\begin{center}
\begin{table}[ddtp]
\caption{Overview of the relevant control protocols for a single
qubit, in the simple case of Hamiltonian suppression. We assume
$H=\hat H+\Delta H,$ where $\hat H$ is the Hamiltonian
estimate. NC stands \vspace{0.2cm}for \vv{Not Corrected}.}

\begin{tabular}{p{5cm}|p{2cm}|p{4cm}|p{3cm}}\label{table1}
\vspace*{-2.5mm}{\bf Protocol}& {\bf Acronym} & $ \hat H$ {\bf corrected by:} 
& $\Delta H$ {\bf corrected by:} \\\hline 
Selective Decoupling & {\tt SelDD} & Decoupling &  NC\\
Maximal Decoupling &  {\tt MaxDD} & Decoupling & Decoupling\\
Feedback-enacted DD &  {\tt FDD} & Feedback & NC \\
Feedback-Enhanced Decoupling &  {\tt FED} & Decoupling & Feedback \\
Decoupling-Enhanced Feedback &  {\tt DEF} & Feedback & Decoupling \\
\end{tabular}
\end{table}%
\end{center}
\end{widetext}

\subsection{Estimation and tuning}
\label{estimation}

The {\tt FED} procedure described above offers a robust solution to
Problem \ref{prob1}. Nevertheless, it also provides us with additional
information we have not used yet: The record of subsequent measurement
outcomes. In fact, the feedback correction we apply needs only the
last outcome, and no extra memory.  However, by recording individual
outcomes, we have (under suitable stationarity assumptions) a way for
estimating $|\epsilon_x|$, i.e., $|\epsilon_x|=\arcsin(\sqrt{
p_1(T_c)})/T_c$.

On the other hand, if a {\tt DEF} protocol is implemented instead,
this allows us information on $|\epsilon_z|$ to be acquired -- and,
with suitable rearrangements, on $|\epsilon_y|$ as well.  We will
illustrate in the next Section how better information about the
underlying Hamiltonian is important to optimize control performance
when ideal conditions are relaxed.

Thus, the acquired information may be exploited to \emph{tune} the
overall control strategy on a better estimate $\hat\sigma$.
Ultimately, this may result into iteratively converging to the correct
Hamiltonian. In turn, this will decrease the need for control actions
in the feedback correction step, as well as the impact of higher-order
errors induced in the concatenated {\tt FED/DEF} protocols for finite
$T_c$.

Attention must be payed to the fact that we are only gathering
information about the {\em absolute value} of $\epsilon_{x,z}$, and to
the fact that we need to estimate the component $\epsilon_y$ in a
second stage, by replacing $\sigma_z$ with $\sigma_y$ operations in
the relevant DD protocol.  In what follows, for simplicity we will not
consider errors in the $y$ direction.  The main steps of a strategy
for Hamiltonian estimation may then be sketched as follows:
    
$(i)$ Run the {\tt FED} protocol $M$ times. Consider ${\sigma_z}$ as
    estimate and record in $R$ the sum of the results of the
    measurement steps. Compute $p^x_1=R/M$ and the estimation for
    $|\epsilon_x|=\arcsin(\sqrt{p^x_1})/T_c$.
    
$(ii)$ Switch to the {\tt DEF} protocol, and run it $M$ times. Compute
    $p^z_1=R/M$ and the estimation for $\epsilon_z=(\arcsin(\sqrt{
    p^z_1})-1)/T_c$.

$(iii)$ Update the estimate, $\sigma^{new}_z=\sigma_z+
    \eta_z\epsilon_z\sigma_z\pm\eta_x\epsilon_x\sigma_x,$ where
    $0<\eta_z,\eta_x\leq 1$ are two weighting parameters to tune,
    depending on the significance of the estimate for finite $M$. In
    the first iteration, the sign of $\eta_x$ can be chosen
    arbitrarily. 

$(iv)$ Iterate steps $(i)$--$(iii)$, changing the sign of $\eta_x$ if
    the frequency of corrections $p_1^x$ is increasing -- indicating a
    worse estimate.  This shall complement on average the sign
    information missing on $\epsilon_x$.

The above procedure will approach the best approximation of the
coupling Hamiltonian in the $x$-$z$ plane, compatibly with the
estimation errors due to a finite number of trials (in any non ideal
setting, i.e. $\Delta t>0$, $M$ is finite). At a subsequent stage, it
is certainly possible to switch to a {\tt FED} strategy employing
$\sigma_y$ transformations, and refine the estimation in the $x$-$y$
plane also.

\section{Numerical Results}
\label{numerical}

If, theoretically, both maximal and feedback-enhanced DD attain the
desired dynamical averaging, from a practical standpoint it is
critical to assess their performance as the ideal conditions are
relaxed. In particular, we will focus on comparing the response of
different single-qubit control strategies when the distance $\Delta t$
between two ideal pulses remains finite, and we will elucidate the
role of the prior information on the Hamiltonian.

As a state-independent performance indicator for control performance,
we invoke {\em entanglement fidelity} $\mathcal{F}$ \cite{schumacher}.
Borrowing from standard information-theoretic results for single-qubit
quantum channels, the latter will be computed indirectly from the
{average fidelity} or from gate fidelity, see e.g.
\cite{nielsen-chuang,nielsen-averagefidelity,1qbitaveragefidelity,violadfs}.
We will use, in particular, the useful result that determines the
average fidelity for a generic quantum operation $\mathcal{T}$ as:
\begin{equation} 
\bar F(\mathcal{T}) = \frac{1}{2} + \frac{1}
{12} \sum_{ j=x,y,z} \tr \left(\sigma_j \mathcal{T}(\sigma_j
)\right)\, .
\end{equation}
Then, $\mathcal{F}$ may be explicitly computed as
$$\mathcal{F}=\frac{(d+1)\bar {F}-1}{d}\,,$$ where $d=d_{\cal S}=2$ in
our case.  Additionally, if ${\cal T}$ is unitary, ${\cal T}(\sigma_j)
=U \sigma_j U^\dagger$, ${\cal F}(T)$ simply reduces to
\cite{violadfs}
\begin{equation}
{\cal F}(T)= \frac{1}{d^2} \left| \tr (U(T)) \right|^2\,.
\label{freeF}\end{equation}

In order to ease the comparison, let us introduce some additional
compact notation for the various strategies under investigation.  We
will consider two {\em selective} strategies, i.e. strategies that are
meant to remove only the estimated Hamiltonian component: The
selective, symmetrized two-pulses Carr-Purcell sequence described in
Section \ref{setting} ({\tt SelDD}), and the {\tt FDD} strategy
introduced and discussed in Section \ref{theory}.  Next, we will
compare some {\em maximal} DD schemes, i.e. strategies that in the
ideal limit $\Delta t\rightarrow 0$ remove any Hamiltonian component
of the evolution: The maximal, four-pulses open-loop DD sequence
sketched in Section \ref{setting} ({\tt MaxDD}), and the two
concatenated strategies presented in Section \ref{application}, the
{\tt FED} and {\tt DEF} protocols.
 
Before discussing in detail the numerical results, it is worth
anticipating that a clear asymmetry emerges in the observed behavior
with respect to errors on the estimated Hamiltonian. For example, the
{\tt FED} and {\tt DEF} performance changes if the faulty Hamiltonian
employed in the simulations is $\sigma_z+\varepsilon\sigma_x$ or
$\sigma_z+\varepsilon\sigma_y$, respectively ($\sigma_z$ representing
the estimated Hamiltonian, as in the previous Section). In a similar
fashion, different, but still ideally equivalent pulse sequences for
{\tt MaxDD}, result in non uniform performance with respect to a given
Hamiltonian.

Higher-order errors emerging from the Magnus series are responsible
for such asymmetry.  If $H$ is the Hamiltonian we wish to suppress
with a sequence $\{G_i\}_{i=1}^{n_g}$ of unitary control actions,
define $\tilde{H}_i=G_i^\dag H G_i$.  Then, recalling
Eq.~\eqref{Hsecond}, the first order correction to the average
Hamiltonian in \eqref{DD} rewrites as
\begin{eqnarray}\label{correction}
\bar H^{(1)}&=&-\frac{i}{2 T_c} \sum_{i>j} \, [\tilde H_i, \tilde
H_j]\Delta t^2 .\end{eqnarray} 
\noindent
Thus, it is straightforward to see that, for instance, for a
Hamiltonian $H=\sigma_z+\varepsilon H_1$, with $\tr(\sigma_z H_1)=0,$
the control sequence $\{I,\sigma_x,\sigma_z,\sigma_y\}$ in
Eq. \eqref{correction} gives non-zero terms only to second order in
$\varepsilon$,
\begin{eqnarray}
\bar H^{(1)}&=&\frac{i \Delta t^2}{T_c} \varepsilon^2 
[\sigma_x H_1 \sigma_x, H_1] \,,\end{eqnarray} \noindent 
whereas the sequence
$\{ I,\sigma_y,\sigma_x,\sigma_z \}$, gives non-trivial terms already
to first-order in $\varepsilon$:
\begin{eqnarray}
\bar H^{(1)}&=&\frac{2 i \Delta t^2}{T_c} \varepsilon [\sigma_y H_1
\sigma_y +H_1, \sigma_x] + {o}(\varepsilon^2)\,.\end{eqnarray} Note
that in both cases the corrections are zero for $\varepsilon=0$,
showing how, could we know the exact Hamiltonian and choose the pulses
freely, we could optimize control performance. Thus, even simple DD
strategies may significantly benefit, in general, from a better
knowledge of the Hamiltonian.  (In fact, accurate knowledge of the
underlying Hamiltonian is one of the key elements for high-level DD
design, as exemplified by a large variety of multipulse DD sequences
in NMR \cite{freeman}).

On the other hand, for a selective DD scheme based e.g. on
$\{I,\sigma_x\}$ and the same Hamiltonian $H=\sigma_z+\varepsilon H_1$
as above, to first order in $\varepsilon$ we obtain
\begin{eqnarray}
\bar H^{(1)}&=&-\frac{i\varepsilon\Delta t^2}{2 T_c} [ \sigma_x H_1
\sigma_x+ H_1, \sigma_z]\,,
\end{eqnarray} 
which is proportional to $\sigma_y$.  This shows how, in order to
retain only the $z$-component of the Hamiltonian with a
$\{I,\sigma_z\}$ scheme, the corrections due to a non-zero $\Delta t$
will be along $y$.  This means that a feedback strategy employing
$C_{\sigma_x}^\Ai$ would correct them along with the $z$-component of
the Hamiltonian, while the use of $C_{\sigma_y}^\Ai$ would not. Such a
perturbative analysis is consistent with the results of the
simulation, and accounts for the observed asymmetries of both the
maximal DD and feedback based strategies. These observations offer
useful insights on the best choice of control actions once the main
component of the Hamiltonian is known.

\begin{figure}[t]
\begin{center}
\includegraphics[width=9cm]{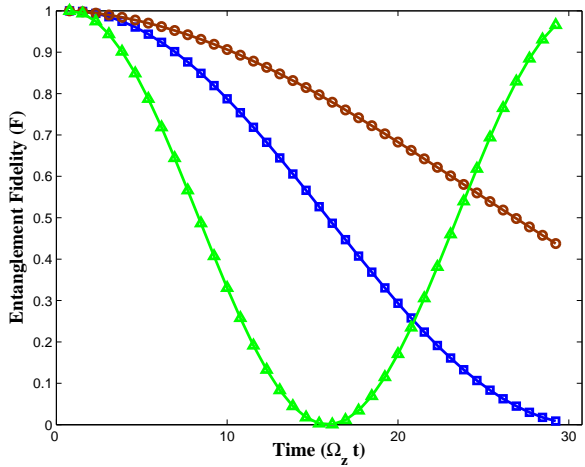}
\caption{{\bf Comparison between selective DD protocols:} (color
online) {\tt FDD} (brown, circles) and {\tt SelDD}, with $\sigma_x$
(blue, squares) or $\sigma_y$ (green, triangles) pulses. Symbols are
displayed every four decoupling cycles to improve readability. Brown
and blue curves correspond to oscillations of the form ${\cal
F}(T)=\cos(\varepsilon_{x,y} T)^2$, respectively -- as following from
Eq.~(\ref{freeF}) for the selectively decoupled evolution.  Actual
Hamiltonian: $H=\Omega_z \sigma_z+ \varepsilon_x
\sigma_x+\varepsilon_y \sigma_y,$ $\varepsilon_x=\Omega_z/20,
\varepsilon_y=\Omega_z/10.$ The estimated Hamiltonian is proportional
to $\sigma_z$. $\Omega T=30$, $\Omega_z \Delta t= 0.04.$}
\label{figsel}
\end{center}
\end{figure}

We now proceed to illustrate and discuss some representative numerical
results, for a generic qubit Hamiltonian of the form $H=\Omega_z
\sigma_z+ \varepsilon_x \sigma_x+\varepsilon_y \sigma_y$.  First, we
compare the {\tt SelDD} with the {\tt FDD} protocol in the presence of
a finite interpulse separation $\Delta t$ (which is the control
non-ideality we focus on).  From Eq.~(\ref{freeF}), the free evolution
corresponds to entanglement fidelity
$${\cal F}(T)=\cos(\Omega T)^2\,,\;\; \Omega \approx \Omega_z\,.$$ A
typical behavior is depicted in Figure \ref{figsel}.  The changes in
the oscillation frequency, $\Omega \mapsto \varepsilon_x$, $\Omega
\mapsto \varepsilon_y$, associated with each {\tt SelDD} protocols,
are clearly seen.  Not only does {\tt FDD} display better performance
for comparable, fixed $\Delta t$, but it also turns out to be less
sensitive to errors in the estimate of the Hamiltonian. In fact, the
{\tt FDD} protocol exhibits better performance as the norm of the
estimation error grows.

In comparing maximal DD strategies, we made sure to employ in each
case the appropriate choice of control pulses, as following from the
analysis of the higher-order corrections presented above.  Different
control scenarios were investigated.  The main features emerging from
our study may be summarized as follows.

As the parameter $\Delta t$ grows within the parameter range
considered, all protocols move away from the ideal behavior in a
regular fashion: Overall, the best performance is attained by the {\tt
FED} strategy, followed by {\tt MaxDD}.  The {\tt DEF} strategy, where
we use the feedback loop to correct the main, estimated Hamiltonian,
shows a sensible reduction of the performance for longer $\Delta t$,
making {\tt MaxDD} more effective in the long time regime.
Independent tests confirmed that this implementation is more sensitive
to the $\Delta t$ parameter increasing. For sufficiently small $\Delta
t$, {\tt DEF} can outperform the {\tt maxDD}, however, compared to
{\tt FED}, it looses its advantage soon.

For a fixed value of both $\Delta t$ and $T$, we also investigated the
performance of the above protocols when the norm of the {estimation
error}{ on the Hamiltonian} is increased.  The same comparative
results are found, with fairly uniform behavior, and no particular
crossings.

\begin{figure}[t]
\begin{center}
\includegraphics[width=9cm]{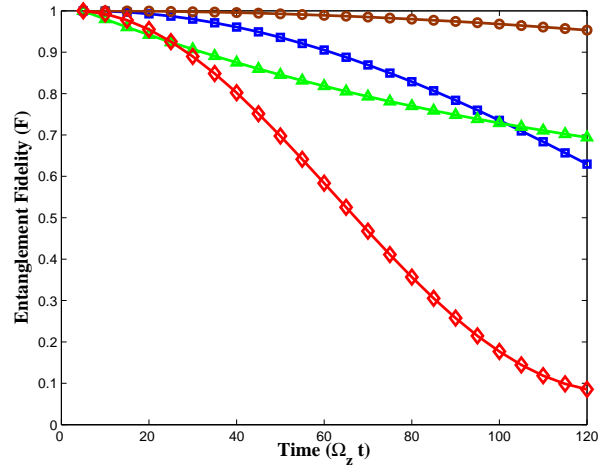}
\caption{{\bf Comparison between maximal DD protocols:} (color online)
{\tt MaxDD} (blue, squares), symmetrized {\tt FED} (brown, circles),
non-symmetrized {\tt FED} (green, triangles), {\tt DEF} (red,
diamonds). Symbols are displayed every two decoupling cycles to
improve readability. Actual Hamiltonian: $H=\Omega_z \sigma_z+
\varepsilon_x \sigma_x+\varepsilon_y \sigma_y,$ $\varepsilon_x=
\varepsilon_y=\Omega_z/10.$ The estimated Hamiltonian is proportional
to $\sigma_z.$ $\Omega_z T=120, \Omega_z \Delta t = 0.32.$}
\label{figmax}
\end{center}
\end{figure}

Based on the above analysis, the most robust and information-efficient
strategy turns out to be the symmetrized {\tt FED} protocol, that is,
the feedback-corrected version of a symmetric selective DD.  It is
worth to notice that the symmetrization on nested {\tt SelDD} is
crucial in determining such advantage: The non-symmetrized version
does not succeed, in general, at outperforming the open-loop strategy.
We summarize in Figure \ref{figmax} the typical behavior of the
protocols under considerations for a fixed $\Delta t$ value. We show
also the simulation results for a {\tt FED} strategy with a
non-symmetrized nested {\tt SelDD}, clearly supporting the above claim
on the importance of symmetrization.

\section{Discussion and Conclusion}

We have designed and characterized different quantum procedures for
exploiting {\em single-bit classical feedback} as a tool for quantum
dynamical averaging -- applicable {\em alone} or {\em in conjunction}
with traditional (deterministic) open-loop decoupling methods.  Beside
providing general (linear-algebraic) necessary and sufficient
conditions for a two-step averaging processes to be implementable in
principle via single-bit feedback on an arbitrary finite-dimensional
open quantum system, our analysis points to novel possibilities for
successfully merging open- and closed-loop techniques toward achieving
robust quantum dynamical control. In particular, the concatenated
feedback-enhanced decoupling protocol proposed and discussed for a
two-dimensional target system in
Sections~\ref{application}-\ref{numerical} is found to exhibit better
performance than its open-loop counterpart, the best possible maximal
DD scheme with respect to the estimated interaction. For higher
dimensional systems, the integration of discrete-time feedback with
open-loop decoupling schemes appears more delicate, as the constraints
imposed by Theorem \ref{thm2} become more demanding. While it is still
conceivable in principle to employ feedback to enhance the performance
of selective open-loop strategies, especially whenever uncertainty is
limited or it affects the free evolution in some structured fashion,
further investigation is needed to address the relevant issues in
detail.

From an information-theoretic perspective, our results offer some
suggestive insights on the role played by {\em classical information}
in various quantum control schemes.  While different schemes become
essentially equivalent and insensitive to system's parameters in the
ideal limit of arbitrarily fast control, {prior} information on the
underlying Hamiltonian plays a central role in the optimization of
different protocols as soon as parameter uncertainties become
important and non-ideal control settings are considered -- as it is
unavoidably the case in practice.  Thus, the role of information in
control design and synthesis cannot be underestimated.  Interestingly,
even if purely open-loop maximal decoupling schemes may be optimized
by using the initial estimation, they do not allow any improvement in
the estimate itself. Thus, the {\em full} power of a feedback strategy
might is expected to appear in {\em adaptive} control settings, like
the simple Hamiltonian estimation setting presented in Section
\ref{estimation}.

From yet another standpoint, the proposed feedback-concatenated
protocols further support the relevance of {\em hybrid} strategies for
robust quantum dynamical control \cite{byrd}. In particular, our
results take an additional step toward establishing the potential of
dynamical decoupling methods for integration with existing passive and
active quantum control techniques -- following the demonstration of
concatenated quantum error correction and decoupling codes
\cite{nicolasDD}, the development \cite{wu,byrd2,viola02} and
implementation of decoherence-free encoded dynamical decoupling
schemes \cite{violadfs}, and the recent development of fault-tolerant,
concatenated dynamical decoupling protocols \cite{lidar-concatenated}.
Beside, as mentioned earlier, generalizations of the proposed
framework to higher-dimensional systems and ancillas, additional
extensions to include non-ideal measurement and realistic control
capabilities are certainly necessary for a more complete picture and
possible contact with experiment to be established.  At the formal
level, we expect that linear-algebraic tools may continue to prove
useful in that respect, along with general methods from quantum
fault-tolerance theory.  In practice, although meeting all the
relevant control requirements appears demanding by present
capabilities, the kind of operations needed for the proposed hybrid
feedback-decoupling protocols have already been implemented separately
in different device settings \cite{expreview} (see also discussion in
\cite{lloyd-uqi}), and experimental progress is steady.  In this
respect, it might be especially intriguing and timely to explore
possible proof-of-principle applications in scalable devices based on
trapped ions, where quantum error correction has been recently
demonstrated \cite{qec-ion} and different isotopic species could be
used in principle for system and ancilla degrees of freedom
\cite{Blinov}, or semiconductor quantum dots, where suggestive
proposals for controlling a slowly fluctuating mesoscopic spin
environment via a suitable probe spin already exist \cite{Taylor}.

Lastly, the analysis of higher order corrections, as carried out in
Section \ref{numerical}, may be seen as a simple instance of the {\em
sensitivity minimization problem}, which is one of the fundamental
problems of robust control theory. In this perspective, it is our hope
that the present work will stimulate further exploration and input
from classical control theory, as well as the quantum control and
system engineering communities.

\vfill
\begin{acknowledgments}
It is a pleasure to thank Lea F. Santos for discussions and a careful
reading of the manuscript.  F. T. acknowledges support from the
ministry of higher education of Italy (MIUR), under project {\em
Identification and Control of Industrial Systems}, from a {{\em Aldo
Gini Fellowship}} for research and studies abroad, and hospitality
from the Physics and Astronomy Department at Dartmouth College --
where this work was performed.  Partial support from Constance and
Walter Burke through their Special Projects Fund in Quantum
Information Science is also gratefully acknowledged.
\end{acknowledgments}

\bibliographystyle{apsrev}
\bibliography{bibliography}

\begin{thebibliography}{47}
\bibitem[{foo()}]{footnote}
\bibinfo{note}{In the case of a two-dimensional target system, an alternative
  control strategy may in principle be considered: that is, to use an initial
  {\tt swap} gate to transfer the initial system state in the decoherence-free
  ancilla subsystem, and restore it when needed with a second {\tt swap}
  operation. While this possibility is definitely viable theoretically,
  feasibility as compared to the proposed scheme may depend in practice on the
  details of the available systems and control capabilities. More importantly
  to our current purposes, it is not clear under what conditions (if any) such
  a {\tt swap} strategy could extend to higher-dimensional control systems.}
\end{thebibliography}

\end{document}